\documentstyle[aps,prb,multicol,psfig]{revtex}

\def\top#1{\vskip #1\begin{picture}(290,80)(80,500)\thinlines \put(
65,500){\line( 1, 0){255}}\put(320,500){\line( 0, 1){5}}\end{picture}}
\def\bottom#1{\vskip #1\begin{picture}(290,80)(80,500)\thinlines \put(
330,500){\line( 1, 0){255}}\put(330,500){\line( 0, -1){5}}\end{picture}}

\begin{document}
\title{{\Large {\bf Coulomb drag between a metal and a Wigner crystal.}}}
\author{V. Braude and A. Stern}
\address{Department of Condensed Matter Physics, The Weizmann Institute of 
Science, Rehovot 76100, Israel}
\date{\today}
\maketitle
\begin{abstract}
We calculated the Coulomb drag contribution to the resistivity and the
transresistivity for a double layer system, in which the passive layer is
a Wigner crystal pinned by impurities, and the active one is a metal. We found
that in quasi one dimensional systems both quantities are suppressed at low
temperatures by the constraints of energy and momentum conservation in
electron-phonon scattering.
In two dimensions both quantities are $\propto T^4$, which is consistent with 
Bloch
law for the contribution of electron-phonon interaction  to the resistivity.
Strong impurities in the Wigner crystal decrease the transresistivity 
significantly
when their number reaches the logarithm of the number of sites in the crystal.
In contrast, 
the drag contribution to the resistivity is found to be  independent of the
impurities, at least as long as their density is kept zero in the
thermodynamic limit.


\end{abstract}

\begin{multicols}{2}
\section{Introduction}

Coupled quantum wells are useful tools for studying electron-electron
interaction in low-dimensional systems both experimentally and
theoretically. By changing the distance between the layers, one can allow or
prohibit interlayer tunneling and also control the interaction between
electrons in different layers. Moreover, by changing parameters such as the
density of electrons, the temperature and the magnetic field one can examine
different electronic phases.

One effect, which has attracted much attention in recent years, is Coulomb
drag \cite{gramila91,gramila93,gramila94,sivan92},
where current $I_{1}$ in one layer (''the active layer'') induces a voltage $%
V_{2}$ in the other (''passive'') layer, which is kept in a zero current
state ($I_{2}=0$). The measured quantity is the drag resistivity, or
trans-resistivity, which for a square sample is defined as $\rho _{D}\equiv -%
\frac{V_{2}}{I_{1}}.$ Since electron-electron collisions have only indirect
consequences for transport properties in a single quantum well, because they
conserve momentum, the Coulomb drag is unique in directly measuring
effects of electron-electron interaction through a transport measurement.

Coulomb drag has been studied extensively in the regime in which carriers in
both layers form Fermi liquids. For low enough density the two-dimensional
electron liquid should condense into a Wigner crystal \cite{bonsall77}. A
clean Wigner crystal should have no electrical resistivity. However, it is
well appreciated by now, from analogous behavior in charge density wave
systems \cite{gruner88}, that an arbitrarily small disorder potential should
pin the crystal so that there are no charge carriers that can flow in
response to an arbitrarily small electric field. The disorder also causes
the crystal to deform, so that it has a finite correlation length, or domain
size. As a result of pinning, the linear response conductivity vanishes.

In this work we study Coulomb drag in a double layer system in which
carriers in the passive layer form a pinned Wigner crystal. The pinning
of the crystal is due to its interaction with strong impurities, whose
density is assumed to vanish in the thermodynamic limit. The common
methods for calculating drag resistivity are not directly applicable to this
case: the Boltzmann equation \cite{jauho93} can be used to calculate the
flow of momentum from the active to the passive layer, but does not
distinguish between momentum flow to the electrons in the Wigner crystal and
momentum flow to the impurities that pin the crystal.  This distinction is
important for separating between two different measurable quantities:  the 
contribution of the passive layer
to the resistivity of the active layer and the trans-resistivity, proportional
to the voltage developing on the
passive layer. The Kubo formalism \cite{zheng93,kamenev95,flensberg95} is 
useful for calculating the trans{\it conductivity}.
However, the inversion of the conductivity matrix, needed to obtain the trans%
{\it resistivity}, is delicate, due to the vanishing zero temperature
conductivity of the passive layer. Thus, this problem necessitates a
different method. 

To that end, we introduce an harmonic potential of frequency $\omega_0$, that 
confines the passive layer and does not allow it to move.
When momentum flows from the active to the passive layer, the passive layer
is displaced from its equilibrium position. Its displacement, which we call 
${\bf u}_{0}$, is
determined by the balance between the momentum flux from the active layer
and the restoring force exerted by the harmonic confining potential. We
calculate ${\bf u}_{0}$, from which we deduce the restoring force and the
transresistivity. The introduction of the harmonic potential into the
problem allows us to calculate the transresistivity directly by using
standard linear response formalism. 

\section{The model and the problem}

The model is defined as follows. There are two conductors (wires in the
quasi- 1D case and layers in the 2D case). Current flows in the active one,
which is in a weakly disordered Fermi liquid state. No current flows in the
passive conductor, which is in a Wigner crystal state pinned by impurities.

The Hamiltonian of the system is given by 
\begin{equation}
H=H_{1}+H_{2}+H_{1-2}.
\end{equation}
Here $H_{1}$ and $H_{2}$ are the Hamiltonians of the active and passive
conductors, and $H_{1-2}$ is the interaction between them.

Thus 
\[
H_1=\sum_{\bf p} \epsilon_{\bf p} c_{\bf p}^\dagger c_{\bf p} + H_{disorder},
\]
where $H_{disorder}$ is the standard interaction of electrons in a metal
with dilute impurities;
the Hamiltonian of the passive conductor, $H_{2}$, describes a crystal
interacting with impurities: 
\begin{eqnarray}  \label{eq:contH}
H_{2}&=&\int \frac{d{\bf x}}{s_0} {\Bigg (} \frac{{\bf p}({\bf x})^2}{2m_2}
\nonumber \\  &&
+\frac{1}{2 }m_2 v_s^2(\nabla_i {\bf u}_j({\bf x}))(\nabla_i 
{\bf u}_j({\bf x})) 
+\frac{1}{2} m_2 \omega_0^2 {\bf u}({\bf x})^2 
\nonumber \\ &&
 -\sum_{imp}\frac{1}{2} A_{G}G^2 s_0
{\bf u}({\bf x})^2 \delta ({\bf x}-{\bf R}_{imp}) {\Bigg )}  , 
\end{eqnarray}
where ${\bf u}$ is the displacement field, ${\bf p}$ - its conjugate
momentum, $s_{0}$ is the size of a unit cell of the Wigner crystal, $m_{2}$
- the mass of electrons constituting the Wigner crystal and $v_s$ - the
sound velocity of the crystal (for simplicity we assume it to be equal for
the longitudinal and transverse modes). $G$ is a basis
vector of the reciprocal Wigner lattice, $A_{G}$ - a parameter, defining the
strength of the impurities in the crystal. We expect the induced field
(which enters into the definition of trans-resistivity) to be independent of $%
\omega _{0}$, and at the end of the calculation $\omega_0$ is taken to zero. 
In our analysis of the Hamiltonian (\ref{eq:contH}) we assume that the 
impurities in the passive layer are dilute and strong, in a sense that is 
defined below. 

The derivation of this Hamiltonian is similar to one given in \cite
{fu-lee78,glaz}.
We start from a discrete harmonic lattice, interacting with impurities: 
\begin{eqnarray}
H_{2} &=& \sum_i\frac{{\bf p}({\bf x}_{i})^2}{2 m_2}+\frac{1}{2}
\sum_{i \neq j}
{\bf u}({\bf x}_i){\bf D}({\bf x}_i-{\bf x}_j){\bf u}({\bf x}_j)
\nonumber \\ & &
+\sum_{i,imp}V({\bf x}_i+{\bf u}({\bf x}_i)-{\bf R}_{imp}).
\end{eqnarray}
Here ${\bf {D}}$ is the dynamical matrix of the crystal, $V$ - the
impurity potential and the third sum extends over lattice and impurity sites. 
Now we write 
\[
{\bf u}={\bf u}^{0}+{\bf u}^{1},
\]
where ${\bf u}^{0}$ gives the static deformation of the crystal due to the
impurities, and ${\bf u}^{1}$ - fluctuations around the new equilibrium.
Expanding everything in small quantities ${\bf u}^1$ to second order, we
get: 
\end{multicols}%
\top{-3cm}
\begin{eqnarray}   \label{eq:discrH}
H_{2} &=&\sum_{i}\frac{{\bf p}({\bf x}_{i})^{2}}{2m_{2}}+\frac{1}{2}%
\sum_{i\neq j}{\bf u}^{0}({\bf x}_{i}){\bf {D}}({\bf x}_{i}-{\bf x}_{j})
{\bf u}^{0}({\bf x}_{j})+%
\sum_{i,imp}V({\bf x}_{i}+{\bf u}^{0}({\bf x}_{i})-{\bf R}_{imp}) \nonumber \\
&&+\frac{1}{2}\sum_{i\neq j}{\bf u}^{1}({\bf x}_{i}){\bf {D}}%
({\bf x}_{i}-{\bf x}_{j}){\bf u}^{1}({\bf x}_{j})+\sum_{i,imp}\frac{1}{2}
({\bf u}^{1}({\bf x}_{i})\nabla
)({\bf u}^{1}({\bf x}_{i})\nabla )V({\bf x}_{i}+{\bf u}^{0}({\bf x}_{i})-
{\bf R}_{imp}),
\end{eqnarray}
\bottom{-3cm}
\begin{multicols}{2}
\noindent where there is no linear in ${\bf u}^{1}$ term, because ${\bf u}^{0}$
minimizes the potential energy, including the impurities, and the crystal is
harmonically pinned at each impurity site. We now assume that the impurity
strength is much larger than elastic forces in the crystal, so that the
lattice adjusts itself in order to minimize the interaction energy with
impurities at each impurity site. This is a strong pinning condition \cite
{fu-lee78}. It is valid when impurities are strong and dilute. 
 As derived in detail below, the impurities are strong when 
\begin{equation}
G^{2}A_{G}>>2m_{2}v_{s}\omega _{0}/a_{0}
\end{equation}
in quasi 1D, and
\begin{equation}
G^{2}A_{G}>>\frac{2\pi m_{2}v_{s}^{2}}{s_{0}\ln (\frac{v_{s}}{a_{0}\omega
_{0}})}
\end{equation}
in 2D. Here $a_0$ is the lattice spacing.

 Assuming strong pinning, ${\bf u}^{0}$ can be found simply by minimizing $%
\sum_{i}V({\bf x}_{i}+{\bf u}^{0}({\bf x}_{i})-{\bf R}_{imp})$ at each
impurity site. Assuming now that the range of the impurities is much larger
than the lattice spacing, but much smaller than the inter-impurity distance, 
we can write 
\[
 \sum_{i}V({\bf x}_{i}+{\bf u}^{0}({\bf x}_{i})-{\bf R}_{imp})\approx
 \sum_{i}V({\bf x}_{i}+{\bf u}^{0}({\bf R}_{imp})-{\bf R}_{imp}),
\]
which is periodic in ${\bf u}^{0}({\bf R}_{imp})$ with lattice period.
Expanding in a Fourier series and taking only the lowest Fourier
components, we may approximate 
\begin{eqnarray}
&&\sum_{i}V({\bf x}_{i}+{\bf u}^{0}({\bf x}_{i})-{\bf R}_{imp})
\nonumber \\ && \quad \quad \approx \sum_{\bf G} 
A_{{\bf G}%
}\cos ({\bf G}({\bf u}^{0}({\bf R}_{imp})-{\bf R}_{imp})),
\end{eqnarray}
which for attractive impurities($A_G<0$) is minimized when $\cos ({\bf G}({\bf u}%
^{0}({\bf R}_{imp})-{\bf R}_{imp}))=1$. Substituting this approximation
in the Hamiltonian (\ref{eq:discrH}), we get: 
\begin{eqnarray}
&&\sum_{i,imp} \frac{1}{2}({\bf u}^{1}({\bf x}_{i})\nabla )({\bf u}^{1}
({\bf x}%
_{i})\nabla )V({\bf x}_{i}+{\bf u}^{0}({\bf x}_{i})-{\bf R}_{imp}) 
\nonumber \\ && \quad \quad
\approx  -\frac{1}{2}A_{{\bf G}}\sum_{i,imp}({\bf u}^{1}({\bf x}_{i})
{\bf G})({\bf u}^{1}(%
{\bf x}_{i}){\bf G})\delta _{{\bf x}_{i}-{\bf R}_{imp}}.
\end{eqnarray}
We now omit the non-dynamical parts of  (\ref{eq:discrH}) (i.e., the parts
that depend only on ${\bf u}^{0}$), add a term $\frac{1}{2}\sum m_{2}\omega
_{0}^{2}({\bf u}({\bf x}))^{2}$ to represent a harmonic confining potential
for the passive layer, and go to the continuum limit. We obtain the
Hamiltonian (\ref{eq:contH}). Since this Hamiltonian does not depend on $%
{\bf u}^{0}$, we can simplify the notation by omitting the superscript from $%
{\bf u}^{1}$ and calling it ${\bf u}$ . It should be remembered, however, that
from now on $\bf u$ is not the deviation of the electron position from a 
lattice site, but its deviation from the position it holds in equilibrium 
(which is shifted from the lattice site by the impurities). 

The interaction between the conductors is given by 
\begin{eqnarray}
H_{1-2}&=&\Bigg( \delta _{{\bf q}+{\bf G}}+\imath \delta _{{\bf k}+
{\bf G}-{\bf q}} {\bf q}{\bf u}_{{\bf k}}-
\nonumber \\ && \, \, \,
\frac{1}{2}\delta _{{\bf k_{1}}+{\bf k_{2}}+{\bf G}%
-{\bf q}}({\bf q}{\bf u}_{{\bf k}_1})({\bf q}{\bf u}_{{\bf k}_2})\Bigg) 
n_{2}U_{{\bf q}%
}c_{{\bf p}+{\bf q}}^{\dagger }c_{{\bf p}},
\end{eqnarray}
where $n_{2}$ is the density of lattice sites in the Wigner crystal. This is
just the inter-layer Coulomb interaction expanded to second order in terms
of the crystal displacement field. We omitted here a term arising from the
static deformation of the Wigner crystal due to impurities. $U_{{\bf q}}$ is 
the
Fourier component of the screened interlayer Coulomb potential. We take it
to be constant at wave length smaller than the interlayer spacing, while at
larger wavelengths it decreases exponentially.

The problem is posed as follows: given a current density ${\bf j}_{1}$ in
the active conductor and ${\bf j}_{2}=0$ in the passive conductor, what are
the electric fields ${\bf E}_{1}+\delta {\bf E}_{1}$ and ${\bf E}_{2}$ in
the conductors? (Here ${\bf E}_{1}$ is the field which would be in the
active conductor in the absence of the passive one). Using the resistivity
matrix \cite{zheng93}, we can write this as 
\begin{equation}
\left( \matrix {{\bf E}_1+\delta{\bf E}_1 \cr {\bf E}_2 \cr}\right) =\left( \matrix {\rho+\delta
\rho & -\rho_D \cr -\rho_D & \rho_W \cr}\right) \left( \matrix {{\bf j}_1 \cr 0 \cr }%
\right) ,
\end{equation}
where $\rho $ is the resistivity of the active layer in the absence of drag.
What we need to find are $\delta \rho $ and $\rho _{D}$. 

We find the field $E_{2}$ by calculating $u_{0},$ since the two are related
by  
\begin{equation}
E_{2}=\frac{m_{2}\omega _{0}^{2}u_{0}}{e},
\end{equation}
while $\delta {\bf E}_{1}$ is found by calculating $\delta {\bf j}_{1}-$ the
negative contribution to the current in the active conductor at a given
field $E_{1}$ due to the drag. Then, for weak drag, 
\begin{equation}
\delta {\bf j}_{1}\equiv -\delta {\bf E}_{1}/\rho.
\end{equation}
We consider both quasi 1D coupled conductors (by which we mean wires of 
several conduction channels, in which electrons  form either a Fermi liquid or 
a Wigner crystal), 
and two dimensional conductors. We focus mostly on the latter. 

\section{Results}

In quasi 1D conductors the transresistivity and the drag contribution to the
resistivity are suppressed at low temperatures in the limit of
vanishing $\omega _{0}$ and infinite $L$, since in that limit momentum 
and energy
cannot be conserved simultaneously for electron- phonon scattering.

In the 2D case the transresistivity is found to be 
\begin{equation}
\rho _{D}=Z\frac{m_{1}}{m_{2}}\frac{U_{0}^{2}T^{4}}{e^{2}\hbar
^{5}v_{fs}v_{s}^{5}n_{1}}\left( 1-\frac{N_{imp}}{2\pi \ln (L/a_{0})}%
\right), 
\label{rhod2d}
\end{equation}
where $U_{0}$ is the zero wave vector component of the interlayer
interaction, $Z$ - a numerical factor given by 
\[  Z=\frac{2}{\pi^2} \int_0^{\infty} \frac{z^4 dz}{\sinh^2(z)}, \]
$m_{1}$ is the electron mass in the active (metallic) layer, $m_{2}$ is the 
electron mass in the Wigner crystal, $v_{s}$ is the sound velocity in the
Wigner crystal, $v_{fs}\equiv \sqrt{v_{f}^{2}-v_{s}^{2}}$, with $v_{f}$
being the  Fermi velocity in the metallic layer (we assume $v_{f}>v_{s}$ and
that $v_{f}$, $v_{s}$ and $v_{fs}$ are of the same order of magnitude) ; $%
n_{1}$ is the density of electrons, $N_{imp}$ is the number of impurities,
and $L/a_{0}$ - the ratio between the size of the system and that of a unit
cell. The impurities are supposed to be dilute enough, so that a condition $%
N_{imp}\ll \ln (L/a_{0})$ is satisfied.

We see that impurities in the passive conductor decrease the
transresistivity. Due to the force they exert on the crystal, part of the
momentum flux from the active layer is transferred to the impurities. This 
part does not lead to drag. Stretching the approximation to its limit,
we find that the influence of the impurities  becomes 
significant once
their number reaches the logarithm of the number of electrons in the Wigner 
crystal.
The impurity strength does not appear in the final expression as long as it 
is much larger than elastic forces in the
Wigner crystal. The temperature dependence is consistent with the 2D
extension of the Bloch law for the resistance due to electron-phonon
scattering.

The drag contribution to the resistivity in 2D is, in the limit of dilute
impurities,  
\begin{eqnarray}
\delta \rho& =&\frac{n_2}{n_1} \rho _{D}  \left(1-\frac{N_{imp}}{2\pi 
\ln (L/a_0)} \right)^{-1} \nonumber  \\ 
&=&Z \frac{n_2 m_1}{n_1 m_2} \frac{U_0^2 T^4}{e^2 \hbar^5
v_{fs} v_s^5 n_1}. 
\end{eqnarray}
This contribution is not suppressed by the impurity induced factor $\left( 1-%
\frac{N_{imp}}{2\pi \ln (L/a_{0})}\right) .$ The momentum, that flows to the
impurities, does increase the resistivity in the active layer, although it
does not induce voltage on the electrons in the passive layer. In the
absence of impurities, the ratio between the transresistivity and the drag
contribution to the resistivity is $n_{1}/n_{2},$ as required by the  Galilean
invariance. Due to this invariance the two electronic systems can flow at
equal speeds without any friction resulting from their interaction. Thus 
when $j_{1}/j_{2}=n_{1}/n_{2}$ we
must have   
\begin{equation}
\delta E=j_{1}\delta \rho -j_{2}\rho _{D}=0.
\end{equation}

\section{Calculation}

As explained in the introduction, the calculation is done using the linear
response formalism. We first consider the trans-resistivity of the passive
conductor and then the contribution of inter-layer interaction to to the
resistance of the active conductor.

\subsection{The transresistivity}

\begin{figure}[htb]
      \centerline{\psfig{figure=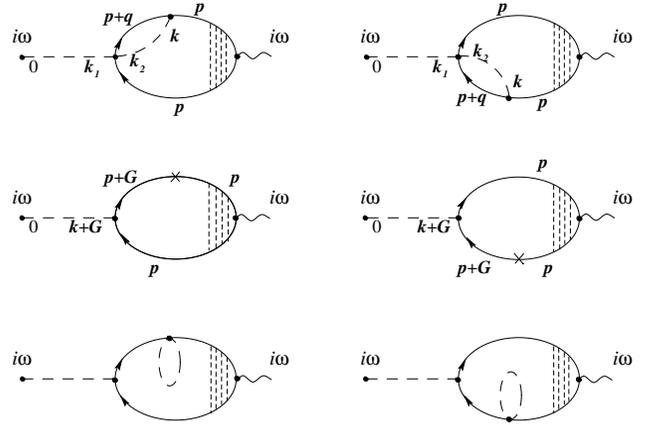,width=8.5cm}}
      \vspace{0.5cm}
\narrowtext
\caption{Diagrams contributing to the drag coefficient to second order in
the electron-phonon potential. The full lines are electron Green's
functions, the long dashed lines are phonon Green's functions, the short
dashed ladder represents a diffuson, and the wavy line is a current vertex. 
The cross represents a periodic
external potential.}
\label{diags}
\end{figure}

From the Kubo formula we find the displacement of the Wigner crystal: 
\begin{equation}
\langle {\bf u}_{0}^{\alpha }\rangle =\lim_{\omega \rightarrow 0}\frac{1}{%
\omega \hbar }\int_{0}^{\infty }dt\exp ^{i\omega t}\langle \lbrack {\bf u}%
_{0}^{\alpha }(t),{\bf j}_{0}^{\beta }]\rangle {\bf E_{1}^{\beta }},
\end{equation}
where ${\bf j}_{0}=e/m\sum_{{\bf q}}{\bf q}c_{{\bf q}}^{\dag }c_{{\bf q}}$ is the
current operator for the active layer and ${\bf E}$ - the applied electric
field (in the same layer). Proceeding as usual \cite{mahan,abrikosov},
we define the correlator $\Pi _{u}$ by 
\begin{equation}
\langle {\bf u}_{0}(\omega )\rangle =\frac{i}{\omega }\Pi _{u}(\omega )\frac{%
e}{m_{1}}{\bf E},
\end{equation}
where we have assumed isotropy. Then in imaginary time the correlator is
given by 
\begin{equation}
\Pi _{u}(i\omega )=\int_{0}^\beta \tau e^{\imath\omega \tau }\langle 
T{{{\bf u}%
}}{}_{0}(\tau ){\bf p}_{0}^{E}\rangle .
\end{equation}
Here $\beta$ is the inverse temperature and ${\bf p}_{0}^{E}$ is the component
of the total momentum of electrons
in the active conductor in the direction of the applied field. This
correlator is then calculated to lowest order of inter-layer interaction. 
The corresponding diagrams are shown in Fig. \ref{diags}. 
Note that the last pair of diagrams vanishes, since the transferred
momentum is zero.

\vspace{0.5cm}

Calculation of the topmost pair of diagrams yields an expression for $\Pi_u$:
\end{multicols}
\top{-3cm}
\begin{eqnarray}
  \Pi_u(\omega \rightarrow 0)&=&\sum_{{\bf k}_1 {\bf k}_2 {\bf p}} -2 i  \left
  (\frac{\hbar}
  {2 m_2} \right)^2 \frac{{\bf q}^2 {\bf q}_x}{\sqrt {\omega_0 \omega_
  {{\bf k}_1} \omega_{{\bf k}_2} 
  \omega_{{\bf k}} }} D_{{\bf k}_1 0}(0) \left( \frac{U_{{\bf q}}}{S} 
  \right)^2 \nonumber \\ &&
  \int \frac{d \epsilon_1}{2 \pi i}
  \int \frac{d \epsilon_2}{2 \pi i} \hbar \omega {\bf \Gamma_p}^x |G_{{\bf p}}
  (\epsilon_1)|^2 
    \beta n_F(\epsilon_1) n_F(-\epsilon_1-\epsilon_2) n_B(\epsilon_2) 
  \left (D_{{\bf k}_2 {\bf k}} (\epsilon_2)-
  D_{{\bf k} {\bf k}_2}^*(\epsilon_2) \right ) \Im  G_{{\bf p}+{\bf q}}
(\epsilon_1+\epsilon_2),
\end{eqnarray}
\widetext
\noindent where in the neglect of Umklapp processes 
${\bf q}={\bf k}= {\bf k}_1+
{\bf k}_2$
and ${\bf k}_1$, ${\bf k}_2,{\bf k}$ are in the Brillouin zone. The frequency 
$\omega_{\bf k}=\sqrt{v_s^2 {\bf k}^2+\omega_0^2}$  is the 
phonon frequency for a clean crystal, $S$ is the area of the sample,
 $\bf {\Gamma}$ is an electron current vertex 
function in the active layer,
$G$ is the conventional electron propagator in the active layer, and finally 
$D$ is the phonon propagator
including impurities, which should be calculated. Substituting for $G$,
which gives delta-functions, integrating over $\epsilon_1$ and $\epsilon_2$, 
and assuming
no Umklapp (since the potential $U_{\bf q}$ is suppressed at large ${\bf q}$),
 so that 
${\bf q}={\bf k}$, we get
\begin{eqnarray}
  \Pi_u(\omega \rightarrow 0)&=& \sum_{{\bf k}_1 {\bf k}_2 {\bf p}} 
  D_{{\bf k}_1 0}(0) \left (\frac{\hbar}
  {2 m_2} \right)^2 \frac{{\bf k}^2 {\bf k}_x U_{\bf k}^2}{S^2\sqrt{\omega_0 
  \omega_{{\bf k}_1} 
  \omega_{{\bf k}_2} \omega_{\bf k} }} \omega \beta n_F(\epsilon_{\bf p}) 
  n_F(-\epsilon_{{\bf p}+{\bf k}})
    \nonumber \\ && \times
   n_B(\epsilon_{{\bf p}+{\bf k}}-\epsilon_{\bf p}) \tau_t \hbar {\bf p}_x 
  \left (D_{{\bf k}2 {\bf k}} ((\epsilon_{{\bf p}+{\bf k}}-\epsilon_{\bf p})/
   \hbar)-
   D_{{\bf k} {\bf k}2}^* ((\epsilon_{{\bf p}+{\bf k}}-\epsilon_{\bf p})/ 
   \hbar) \right),
\end{eqnarray}
\noindent where $\tau_t$ is the electron transport scattering time for the 
active
conductor and $\epsilon_{\bf p}$ - electron energy for the active conductor.
Now we sum over ${\bf p}$, assuming $T \ll \epsilon_F$ and $1/d \ll k_f$ (where
$d$ is the spacing between the conductors and $k_f$ the Fermi wave vector). 
Then

\begin{eqnarray} \label{eq:Dcorrelator}
  \Pi_u(\omega \rightarrow 0)&=&\sum_{{\bf k}_1 {\bf k}_2 } -\frac{1}{2} N(0)
   D_{{\bf k}_1 0}(0) 
  \left (\frac{\hbar}
  {2 m_2} \right)^2 \hbar \omega \beta \nonumber \\  && \times
\frac{{\bf k}^2 {\bf k}_x^2 U_{\bf k}^2}{S^2 
  \sqrt{\omega_0 \omega_{{\bf k}_1} 
  \omega_{{\bf k}_2} \omega_{\bf k} }}   
   \frac{\hbar v_F k \cos \theta}{4 \sinh^2
  (\beta \hbar v_f k \cos \theta /2)}  
 \left (D_{{\bf k}2 {\bf k}} (v_F k \cos 
  \theta)-
   D_{{\bf k} {\bf k}2}^* (v_F k \cos \theta) \right),
\end{eqnarray}
\bottom{-2.7cm}
\begin{multicols}{2}
\noindent where $N(0)$ is the extensive density of states for the metallic 
conductor.
To proceed further, we have to calculate the phonon propagator $D$ in the 
presence of impurities.
Here we calculate the phonon 
propagator in different dimensions, assuming that strength of the 
impurities is much
larger than typical elastic forces in the crystal, i.e., strong pinning.
To formulate quantitatively
 this condition, we consider the static deformation of the Wigner lattice due
  to all
impurities. In our derivation of $H_2$ above, we assumed that the fragment of 
the lattice, that interacts with each impurity, is positioned in a way, that 
minimizes this interaction. In doing so, we neglected elastic forces on the 
scale of inter-impurity distance and found out that, up to a lattice constant,
the displacement of a fragment interacting with an impurity at ${\bf R}_{imp}$
satisfies ${\bf u}^0={\bf R}_{imp}$. The elastic forces between fragments will 
modify this displacement. We now express this modification in terms of phonon 
propagator. In coordinate representation the propagator is
\begin{equation}
  \tilde{D}(\omega, {\bf x}_1, {\bf x}_2) \equiv \int \frac{d t}{\hbar} e^
  {\imath \omega t} \left(-i 
   \theta(t) \langle [u_{{\bf x}_1}(t), u_{{\bf x}_2}(0)]\rangle \right),
\end{equation}
and it can be written in momentum 
representation as: 
\begin{equation}  \label{eq:xpreps}
   \tilde{D}(\omega, {\bf x}_1, {\bf x}_2)= 
      \sum_{{\bf k}_1, {\bf k}_2} \frac{\hbar D_{{\bf k}_1, {\bf k}_2}(\omega)
    e^{\imath {\bf k}_1 {\bf x}_1-\imath {\bf k}_2 {\bf x}_2}}{2 N_2 m_2 \sqrt
    {\omega_{{\bf k}_1} \omega_{{\bf k}_2}}},
\end{equation}
where $N_2$ is the number of sites in the Wigner lattice.
The propagator 
$\tilde{D}$ relates a point force $f$ applied 
at ${\bf x}_2$ to the displacement $u_{{\bf x}_1}$ it generates at ${\bf x}_1$
\begin{equation}
  u_{{\bf x}_1}=-\tilde{D}(\omega =0, {\bf x}_1, {\bf x}_2) f.
\end{equation}
The force exerted on the lattice by an impurity at ${\bf R}_i$ is (assuming 
${\bf G}({\bf u^0}({\bf R}_i)-{\bf R}_i)\ll 1$) 
\[G^2  A_G ({\bf R}_{i}- {\bf u}^0_{{\bf x}_{i}}). \]

Consequently, the displacement ${\bf u}_{{\bf x}_k}^0$ satisfies the equation
\begin{equation}
  {\bf u}^0_{{\bf x}_k}=-\sum_{i}\tilde{D}^{f}(\omega =0, {\bf x}_k, 
  {\bf x}_i)G^2 
  A_G ({\bf R}_{i}-{\bf u}^0_{{\bf x}_{i}}),
\label{balancepinning}
\end{equation}
where 
$\tilde{D}^{f}(\omega =0, {\bf x}_1, {\bf x}_2) $ 
is the propagator of a clean crystal in the coordinate representation.
When the impurities are dilute, 
$-\tilde{D}^{f}(\omega =0, {\bf x}_k, {\bf x}_k)>>
-\tilde{D}^{f}(\omega =0, {\bf x}_k, {\bf x}_i)$ for $i \neq k$.
Under this condition we may retain only the term with $i=k$ in the sum, and the
strong pinning condition, requiring that $u^0>>r_{imp}-u^0$ becomes:
\begin{equation} \label{eq:strong}
  -\tilde{D}^{f}(\omega =0, {\bf x}_k,{\bf x}_k)G^2 A_G>>1.
\end{equation}
Physically, the neglect of the $i\ne k$ terms in the sum (\ref{balancepinning})
implies that the local displacement $u^0_{{\bf x}_k}$ is determined by a 
balance 
between the potential exerted by the closest impurity and the confining 
harmonic potential characterized by $\omega_0$. 

In 1D the condition (\ref{eq:strong}) is 
\begin{equation}
  G^2 A_G>>2 m_2 v_s \omega_0/a_0,
\end{equation}
while in 2D:
\begin{equation}
  G^2 A_G>>\frac{2 \pi m_2 v_s^2}{s_0 \ln(\frac{v_s}{a_0 \omega_0})},
\end{equation} 
where $s_0$ is the area of the unit cell.

The calculation of the phonon propagator in the presence of the impurities goes
as follows. We start from
a differential equation satisfied by the propagator in the coordinate
representation:
\begin{eqnarray}
&&  \left (\omega^2+v_s^2 \nabla^2-\omega_0^2+\sum_i \frac{s_0 G^2 A_G}{m_2}
  \delta({\bf x}-{\bf x}_i) \right) \tilde{D}(\omega, {\bf x}, {\bf x}')
\nonumber \\ && \quad \quad \quad =
\frac{s_0 \delta({\bf x}-{\bf x}')}{m_2},
\end{eqnarray}
where the sum is over the impurities.
This equation can be solved in terms of the free propagator 
\begin{eqnarray} \label{eq:xpropag}
  \tilde{D}(\omega, {\bf x}, {\bf x}')&=&\tilde{D}^f(\omega, {\bf x}-{\bf x}')
-\nonumber \\ && 
   \sum_{i} 
   G^2 A_G \tilde{D}^f(\omega, {\bf x}-{\bf x}_i) \tilde{D}(\omega, {\bf x}_i,
  {\bf x}'),
\end{eqnarray}  
which gives 
\begin{equation}
  \tilde{D}(\omega, {\bf x}_k, {\bf x}')=\sum_{j}({\cal A}^{-1})_{k j} 
  \tilde{D}^f(\omega, 
  {\bf x}_j-{\bf x}'),
\end{equation}
with matrix $\cal A$ defined by
\begin{equation}   \label{eq:matrA}
  {\cal A}_{j i}=\delta_{j i}+G^2 A_G \tilde{D}^f(\omega, {\bf x}_j-{\bf x}_i).
\end{equation}
Substituting this back into (\ref{eq:xpropag}), we obtain
\end{multicols}
\begin{eqnarray}
  \tilde{D}(\omega, {\bf x}, {\bf x}')&=&\tilde{D}^f(\omega, {\bf x}, 
  {\bf x}')-
\sum_{i j}G^2 A_G 
  \tilde{D}^f(\omega, {\bf x}-{\bf x}_i) 
  ({\cal A}^{-1})_{i j} \tilde{D}^f(\omega, {\bf x}_j-{\bf x}').
\end{eqnarray}
\noindent Using (\ref{eq:xpreps}), we can obtain an expression for the 
propagator
in the momentum representation:
\begin{eqnarray}
  D_{{\bf k}_1 {\bf k}_2}(\omega)&=&D^f_{{\bf k}_1}(\omega) \delta_{{\bf k}_1 
  {\bf k}_2}-\sum_{i j}\frac{\hbar}
  {2 N_2 m_2} 
 G^2 A_G ({\cal A}^{-1})_{i j} D^f_{{\bf k}_1}(\omega) D^f_{{\bf k}_2}(\omega) 
  e^{-\imath {\bf k}_1 {\bf x}_i+
  \imath {\bf k}_2 {\bf x}_j }   \nonumber \\
  &=& \frac{2 \omega_{{\bf k}_1} \delta_{{\bf k}_1 {\bf k}_2}}{\hbar ((\omega+
  \imath \delta)^2-\omega_{k_1}^2)}-
   \sum_{i, j} \frac{2 \sqrt{\omega_{{\bf k}_1} \omega_{{\bf k}_2}} G^2 A_G 
   ({\cal A}^{-1})_{i j}}{\hbar 
   m_2 N_2}
  \frac{e^{-\imath {\bf k}_1 {\bf x}_i} e^{\imath {\bf k}_2 {\bf x}_j}}
  {((\omega+
  \imath \delta)^2-\omega_{{\bf k}_1}^2)
  ((\omega+\imath \delta)^2-\omega_{{\bf k}_2}^2)},
\end{eqnarray}
where the free propagator in the momentum representation was used:
\begin{equation}
  D^f_{{\bf k}_1 {\bf k}_2}(\omega)=\delta_{{\bf k}_1 {\bf k}_2} 
D^f_{{\bf k}_1}
  (\omega)=\frac{2 \omega_{{\bf k}_1}}
  {\hbar ((\omega+\imath \delta)^2-\omega_{{\bf k}_1}^2)}.
\end{equation}
In the limit of strong pinning, the propagator satisfies an important 
relationship:
\begin{equation} \label{eq:vanish}
  \sum_{\bf k} \frac{e^{\imath {\bf k}_1 {\bf x}_l} D_{{\bf k}_1 0}(\omega=0)}
  {\sqrt{\omega_0 \omega_{{\bf k}_1}}}
  \propto {\cal O}\left(\frac{1}{G^2A_G{\tilde D}^f(\omega=0,{\bf x}=0)}\right
   )\approx 0,
\end{equation}
\bottom{-3cm}
\begin{multicols}{2}
\noindent where ${\bf x}_l$ is an impurity site. This relationship means 
that a force, when 
applied at an impurity site, causes a
much smaller deformation, than when applied anywhere else, since it is
opposed by the impurity. 
The proof of this relationship follows from the equation satisfied by the 
propagator $\tilde D$, namely ${\tilde D}={\cal A}^{-1}{\tilde D}^f$. In the 
strong pinning limit ${\cal A}_{ij} \approx G^2 A_G \tilde{D}^f(\omega, 
{\bf x}_j-{\bf x}_i)$ and hence (\ref{eq:vanish}).

From the above relationship it follows that for all $\Omega\gg\omega_0$,
\begin{equation}
  \sum_{{\bf k}_1} \frac{D_{{\bf k}_1 0}(0) (D_{{\bf k}_2 {\bf k}}^i
    (\Omega)-(D_{{\bf k} {\bf k}_2}^i(\Omega))^*)}{
     \sqrt{\omega_0 \omega_{{\bf k}_1}}}\approx 0,
\end{equation}
where $D^i\equiv D-D_f$ is a part of the propagator due to the impurities. 
To see this, we note using Eq. (34) that $D_{{\bf k}_1 0}(0)/\sqrt{
\omega_{{\bf k}_1}}$ is a 
sharply peaked function of
${\bf k}_1$ around ${\bf k}_1=0$: its free part is proportional to 
$\delta_{{\bf k}_1, 0}$, and the part due
to impurities is peaked around zero with width $\omega_0/v_s$, because of the 
denominator $1/\omega_{{\bf k}_1}^{3/2}$. In contrast, $D_{{\bf k}_2 {\bf k}}^i
(\Omega)$ varies slowly in that region, 
except for the 
exponential factor $e^{-\imath {\bf k}_2 {\bf x}_i}=e^{-\imath {\bf k} 
{\bf x}_i} e^{\imath {\bf k}_1 {\bf x}_i}$. 
Hence the summation is actually of 
the form (\ref{eq:vanish}).

Using the last relationship in (\ref{eq:Dcorrelator}), and taking into account
 that the typical energy transfer in an inter-layer scattering event, 
$v_Fk\cos{\theta}$, is of the order of the temperature, and thus much larger 
than $\omega_0$, we see that the
effect of impurities enters Eq. (\ref{eq:Dcorrelator}) only in the first 
phonon propagator, where the frequency is zero, while for the
second phonon propagator, that carries a non-zero frequency, only the free 
part of $D$ contributes.  Thus, 
\begin{minipage}{3.2in}
\begin{eqnarray} \label{eq:PIu}
   \Pi_u &=&  -N(0)\sum_{{\bf k}_1 {\bf k}}D_{{\bf k}_1 0}^R(0)(\frac{\hbar}
   {2 m_2})^2  
\frac{k^2 
    U_{\bf k}^2}{S^2 \sqrt{ \omega_0 \omega_{{\bf k}_1} \omega_{{\bf k}_2} 
\omega_
    {\bf k}}} \nonumber \\ && \times \hbar \omega \beta \tau_t {\bf k}_x^2
  \frac{v_f \hbar k \cos \theta}{4 \sinh^2(\beta v_f \hbar k \cos \theta/2)}
   \nonumber \\ && \times
(\sum_\mp \mp \frac{\delta_{{\bf k}_2 {\bf k}}}{\hbar} \delta(v_f k \cos 
   \theta \mp \omega_k)).
\end{eqnarray}
\end{minipage}

To proceed further with the calculation, we should consider different
dimensions.
In 1D (where electrons in the active
wire are in the Fermi liquid regime) $\cos \theta =\pm 1$, so that the momentum
conservation condition is
never satisfied (in the limit of vanishing $\omega _{0}$ and vanishing 
temperature). This prevents the
crystal from getting any momentum from the conductor and thus the
trans-resistivity is zero in the low temperature limit.

The situation is different in two dimensions. 
In order to calculate $\Pi_u$ in 2D we have first to find $D_{0 0}(0)$. For 
this
we write approximately eq.(\ref{eq:matrA}) for $\omega=0$:
\begin{equation}
 {\cal A}_{j i}(\omega=0)=G^2 A_G \tilde{D}^f(\omega=0, {\bf x}_j-{\bf x}_i).
\end{equation}
Now we assume that the number of impurities is so low that the distance
between them is of the order of the system size (hereby neglecting 
fluctuations in which impurities could be close together). Then the matrix
${\cal A}$ satisfies
\begin{equation}
  {\cal A}_{i j}(\omega=0)<<{\cal A}_{i i}(\omega=0)   
\end{equation}
for $i \neq j$. The matrix $A$ then can be inverted approximately:
\begin{eqnarray}
  {\cal A}_{i j}^{-1}(\omega=0) &\approx& S(i, j) \frac{1}{G^2 A_G 
  (\tilde{D}^f(\omega=0, {\bf x}=0))^2}
\nonumber \\ && \times
  \tilde{D}^f(\omega=0, {\bf x}_j-{\bf x}_i),
\end{eqnarray}
where $S(i, j) =1$ for $i=j$ and $-1$ for $i \neq j$.
Substituting this into (\ref{eq:xpropag}), we find $D_{0 0}(0)$ in 2D: 
\begin{equation}  \label{2Dprop}
D_{0 0}(0)=-\frac{2}{\hbar \omega_0}+\frac{4 \pi v_s^2 N_{imp}} {\hbar L^2
\omega_0^3 \ln(\frac{v_s}{a_0 \omega_0})},
\end{equation}
where $N_{imp}$ is the number of impurities. The $\ln(\frac{v_s}{a_0 \omega_0%
})$ originates from the expression for $\tilde{D}_f(\omega=0, {\bf x}_1=0, 
{\bf x}%
_2=0)$ in 2D.

Substituting expression (\ref{2Dprop}) for the propagator into Eq.(\ref
{eq:PIu}) and choosing 
\[
\omega_0 \sim \frac{v_s}{L}, 
\]
we obtain the transresistivity Eq. (\ref{rhod2d}).

As evident in Eq. (\ref{rhod2d}),  the impurities decrease the
trans-resistivity as they absorb part of the momentum transferred from the 
active layer. That part of the transferred momentum does not generate a drag 
voltage, and therefore does not lead to trans-resistivity. Under the strong 
pinning approximation, the
influence of the impurities becomes significant when their number is
comparable to $\ln(L/a_0)$, even though their density is still zero in the
thermodynamic limit.

\subsection{Drag contribution to resistivity}

\begin{figure}[tbp]
\centerline{\psfig{figure=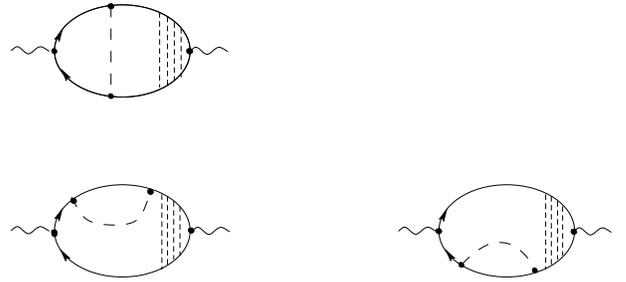,width=8cm}}
\narrowtext
\caption{Diagrams contributing to reduction of conductivity of the active
conductor}
\label{conddiags}
\end{figure}

We now turn to the way electron-electron scattering between different layers 
affects the resistivity of the active layer. We again apply the Kubo formula 
to find, to second order in screened inter-layer interaction,  the current
density in the active layer as a response to an electric field in that same 
layer (we expect it to be negative as the drag should decrease the
current): 
\begin{equation}
{\bf j}^\alpha =Re \frac{1}{\hbar \omega S} \int_0^\infty dt \, e^{\imath 
\omega t} \langle \lbrack \, {\bf j}_0^\alpha (t), {\bf j}_0^\beta \rbrack 
\rangle {\bf E}^\beta.
\end{equation}
Introducing the correlator from 
\begin{equation}
{\bf j}^\alpha= \frac{\imath}{\omega} \lim_{\omega \rightarrow 0} \Im
\Pi_j^{\alpha \beta}(\omega) \, {\bf E}^\beta,
\end{equation}
we have in imaginary time: 
\begin{equation}
\Pi(i\omega)=-\frac{1}{S} \int_0^\beta d\tau \, e^{\imath \omega \tau} \langle
 T {\bf j}_0^x(\tau) {\bf j}_0^x(0) \rangle,
\end{equation}
where we again assumed isotropy. The corresponding diagrams are shown in
Fig. \ref{conddiags}.

Calculating them, we obtain 
\end{multicols}
\top{-3cm}
\widetext
\begin{eqnarray}
\Pi &=& -(\frac{e}{m_1})^2\sum_{{\bf p}, {\bf k}} k^2
\left( \frac{ U_k}{S} \right)^2
n_2 \frac{\hbar}{2 m_2 \omega_k} \omega \beta 
 \frac{n_F(\epsilon_{\bf p})-n_F(-%
\epsilon_{{\bf p}+{\bf k}})}{4 \sinh^2(\beta (\epsilon_{{\bf p}+k}- 
\epsilon_{{\bf p}%
})/2)} \Im D_{{\bf k} {\bf k}}^R(\epsilon_{{\bf p}+k}- \epsilon_{{\bf p}})
\hbar \tau_t^2 \bf{k}_x^2.
\end{eqnarray}
\bottom{-3cm}
\begin{multicols}{2}

In one dimension, when the passive layer is clean, the constraints of 
energy and momentum conservation 
together with the Pauli principle suppress the resistivity at low
temperature. In the presence of impurities in the Wigner crystal 
momentum conservation is relaxed, and we obtain a contribution, which is
inversely proportional to the length of the system. This contribution
can be neglected, since in the macroscopic limit
we assume the length to go to infinity, while the density of impurities
is kept zero.

In two dimensions, in the absence of impurities momentum and energy can 
be conserved
simultaneously, and inter-layer interaction yields the following contribution 
to the current 
\begin{equation}
\delta {\bf j}_1= -\frac{e^2 n_2 U_0^2 T^4 \tau_t^2}{\hbar^5 m_1 m_2
v_{fs} v_s^5} Z {\bf E}_1 \equiv \delta \sigma {\bf E}_1,
\end{equation}
from which the drag contribution to the resistivity is easily found: 
\begin{equation}
\delta \rho = -\rho^2 \delta \sigma= \frac{n_2 U_0^2 T^4 m_1}{e^2 \hbar^5 m_2 
v_{fs} v_s^5
n_1^2} Z.
\label{dragres2d}
\end{equation}
In the presence of impurities the phonon propagator is composed of a free part
 and an impurity-induced part. The contribution of the former remains 
Eq. (\ref{dragres2d}), while the contribution of the latter is  
again inversely proportional to
the system size. Hence in the macroscopic limit the impurities do not 
affect Eq. (\ref{dragres2d}). 

When the passive layer is clean, the ratio between the trans-resistivity 
and the drag contribution to resistivity of the active 
layer is $n_1/n_2$.
This is consistent with the Galilean invariance requirement.

\section{Summary}

We considered a bi-layer system in which the passive layer is a
Wigner crystal pinned by impurities and the active layer is a conductor, and 
calculated the transresistivity between two layers and the drag contribution 
to the
resistivity of the active layer. We focused on the case of a small number of
strong impurities, whose density vanishes in the thermodynamic limit. We found
 that in quasi 1D case both quantities are exponentially small at low 
temperature
due to constraints of energy and momentum conservation.

In two dimensions we found a $T^4$--dependence of both quantities,  consistent
 with the Bloch law for the
electron-phonon interaction contribution to resistivity. Impurities in the
Wigner crystal decrease the transresistivity significantly when their number
reaches the logarithm of the number of sites in the crystal. In the limit of 
strong pinning the transresistivity and the active layer resistivity are 
independent of the impurity strength.

While a minute density of impurities is sufficient to affect the 
transresistivity, we find that it does not significantly affect the 
contribution of the Wigner crystal to the resistivity of the active layer. 

\acknowledgments
V.B. wants to thank Y. Levinson for useful discussions.
We acknowledge support of the Israel Science foundation, the DIP-BMBF 
foundation, the 
Victor Ehrlich chair and the Albert Einstein Minerva center for theoretical
physics at the Weizmann institute. 

\end{multicols}

\begin{references}
\bibitem{gramila91}  T.~J.~Gramila et {\it al.,} Phys.~Rev.~Lett.~ {\bf 66},
1216 (1991).

\bibitem{gramila93}  T.~J.~Gramila et {\it al.,} Phys.~Rev.~B {\bf 47},
12957 (1993).

\bibitem{gramila94}  T.~J.~Gramila et {\it al.,} Physica B {\bf 197},  442
(1994).

\bibitem{sivan92}  U.~Sivan, P.~M.~Solomon and H.~Shtrikman, 
Phys.~Rev.~Lett.~{\bf 68}, 1196 (1992).

\bibitem{gruner88}  G.~Gr\"{u}ner, Rev.~Mod.~Phys. {\bf 60}, 1129 (1988).

\bibitem{jauho93}  A.-P.~Jauho and H.~Smith, Phys.~Rev.~B {\bf 47}, 4420
(1993).

\bibitem{zheng93}   L.~Zheng and A.~H.~MacDonald, Phys.~Rev.~B {\bf 48},
8203 (1993).

\bibitem{kamenev95}   A.~Kamenev and Y.~Oreg, Phys.~Rev.~B {\bf 52}, 7516
(1995).

\bibitem{flensberg95}   K.~Flensberg, B. Y.-K.~Hu, A.-P.~Jauho and
J.~M.~Kinaret,  Phys.~Rev.~B {\bf 52}, 14761 (1995).

\bibitem{fu-lee78}  H.~Fukuyama and P.~Lee, Phys.~Rev.~B{\bf 17}, 535 (1978).

\bibitem{glaz}   L.~I.~Glazman, I.~M.~Ruzin and B.~I.~Shklovskii,
Phys.~Rev.~B {\bf 45}, 8454 (1993).

\bibitem{callaway}   J.~Callaway, {\it Quanntum Theory of Solid State}
(Academic  Press, New York, 1974).

\bibitem{mahan}   G.~D.~Mahan, {\it Many-Particle Physics}, 2nd ed.~(Plenum
Press,  New York, 1990).

\bibitem{abrikosov}  A.~A.~Abrikosov, L.~P.~Gorkov and I.~E.~Dzyaloshinski, 
{\it Methods of Quantum Field Theory in Statistical Physics},  edited by
R.~A.~Silverman (Prentice-Hall, Englewood Cliffs,  New Jersey, 1963).

\bibitem{bonsall77}   L.~Bonsall and A.~A.~Maradudin, Phys.~Rev.~B {\bf 15},
1959 (1977).

\bibitem{Born}   M.~Born and K.~Huang, {\it Dynamical Theory of Crystal
Lattices}  (Oxford U.~P., Oxford, 1954).

\end{references}
\end{document}